\documentstyle[12pt]{article}
\newcommand{\beq}{\begin{equation}}
\newcommand{\eeq}{\end{equation}}
\newcommand{\bea}{\begin{eqnarray}}
\newcommand{\eea}{\end{eqnarray}}

\newcommand{\CH}{{\cal H}}

\newcommand{\pa}{\partial}

\newcommand{\nn}{\nonumber}

\begin{document}
\topmargin 0pt
\oddsidemargin 5mm
\headheight 0pt
\topskip 0mm

\addtolength{\baselineskip}{0.20\baselineskip}

\pagestyle{empty}

\begin{flushright}
OUTP-98-47P\\
19th June 1998\\
hep-th/9806229
\end{flushright}

\begin{center}

\vspace{18pt}
{\Large \bf Beta Function Constraints From Renormalization Group Flows
in Spin Systems}

\vspace{2 truecm}

{\sc Jo\~ao D. Correia\footnote{e-mail: j.correia1@physics.ox.ac.uk}}

\vspace{1 truecm}

{\em Department of Physics, University of Oxford \\
Theoretical Physics,\\
1 Keble Road,\\
 Oxford OX1 3NP, UK\\}

\vspace{3 truecm}

\end{center}

\noindent
{\bf Abstract.} Inspired by previous work on the constraints that
duality imposes on beta functions of spin models, we propose a
consistency condition between those functions and RG flows at
different points in coupling constant space. We show that this
consistency holds for a non self-dual model which admits an exact RG
flow, but that it is violated when the RG flow is only approximate. We
discuss the use of this deviation as a test for the ``goodness'' of
proposed RG flows in complicated models, and the use of the proposed
consistency in suggesting RG equations.

\vfill
\newpage
\setcounter{page}{1}
\pagestyle{plain}

\section{Introduction}

Recent years have seen an explosion of work in dualities of quantum
field theories, and it was to be expected that such work would have
implications for Statistical Mechanics, especially when spin systems
are concerned, since they are very simple avatars of QFTs. 

This has indeed happened, and the work reported below has its basis in the
proposals of Daamgard and Haagensen \cite{D&H}. They established a
relation between the beta function and (exact) dualities of spin models,
then proceeded to apply this relation to obtain a transparent connection
between self-dual points and phase transitions. 

 Here we report on perhaps the most immediate extension of that work: 
instead of considering dualities, we will look at spin systems having an
exact RG flow. A consistency relation for the beta functions (similar to
that of dualities) is then derived, and verified for one such model. We
then address the case of models without exact RG flows, first by
considering an artificial perturbation of an exact RG, and then by
examining an actual case found in the literature. Not surprisingly, the
consistency conditions are no longer obeyed. We also mention the
possibility of using beta function consistency to suggest new RG schemes,
and provide a simple example.  We conclude by discussing possible
applications of the results derived.

\section{Consistency Conditions}

Consider a spin system with coupling constants ${\bf K} =
(K_1,K_2,\ldots,K_N)$, and an exact RG flow which takes the system
from point ${\bf K}$ to point ${\bf K}'$ in coupling constant space in
such a way that the Hamiltonian retains its form. We write
\beq K_i'=t_i(K_1,\ldots,K_N) \label{(1)} \eeq
Defining the beta functions as describing the change of couplings
associated with a change of lattice spacing
\beq \beta(K_i)=-a \frac{\pa K_i}{\pa a} \label{(2)} \eeq
and applying the operator $\pa/\pa a$ to (\ref{(1)}) yields
\bea -a\frac{\pa K_i'}{\pa a} &=& -a \frac{\pa}{\pa a}
t_i(K_1,\ldots,K_N) \nn \\
&=& -a \frac{\pa K_j}{\pa a} \frac{\pa t_i}{\pa k_j} \label{(3)} \eea
where a sum over repeated indices is implied. By identifying the LHS
of (\ref{(3)}) with $\beta(k_i')$ and using (\ref{(2)}) on the RHS, we
arrive at
\beq \beta(K_i') = \beta(K_j) \frac{\pa t_i}{\pa K_j} \label{(4)} \eeq

The system of equations (\ref{(4)}) represents a strong constraint on
the possible forms of the beta functions: below we will see explicitly
that it does not hold for inexact RG transformations. Notice also that
the key assumption made was that the RG transformation mapped the
Hamiltonian $\CH ({\bf K})$ into  $\CH ({\bf K'})$, which is of the
same form as the original one. The duality transformations are a
special case, which supplement equation (\ref{(1)}) with the condition
$t(t({\bf K}))={\bf K}$; systems which possess both dualities and
quasi-exact RG flows are a fertile ground for work on consistency
constraints \cite{sob}.

Before proceeding, it is worth relating the beta functions to the
correlation lengths, which are more readily obtained from real-world
or numerical data. Consider a spin system with coupling tuned such
that ${\bf K} = (K_1,K_2^*,\ldots,K_N^*)$ where the $K_i^*$ are fixed
points of the RG transformation which changes lattice spacing from $a$
to $a+\delta a$. Under such a transformation, the $K_i^*$ are left
unchanged, but $K_1 \rightarrow K_1'$, and the correlation length
transforms as
\beq
\xi(K_1',K_2^*,\ldots,K_N*)=\xi(K_1,K_2^*,\ldots,K_N^*)\left(1+\frac{\delta
a}{a}\right) \label{(5)} \eeq
By expanding the RHS of the previous equation in a Taylor series around
$K_1'$ and using (\ref{(2)}) we find that
\beq \beta(K_1) = -\frac{\xi(K_1,K_2^*,\ldots,K_N^*)}{\frac{\pa
\xi}{\pa K_1} (K_1,K_2^*,\ldots,K_N^*)} \label{(6)} \eeq
Hence, given an RG scheme, the beta functions can be computed if the
correlation lengths are known. Another relation 
between beta functions can be obtained by considering
\beq \xi(K_1',K_2',\ldots,K_N')=\xi(K_1,K_2,\ldots,K_N)\left(1+\frac{\delta
a}{a}\right) \label{(7)} \eeq
and performing an N-variable Taylor expansion keeping only first order
terms, leading to
\beq \beta(K_i) \frac{\pa \xi}{\pa K_i} (K_1,K_2,\ldots,K_N) = - \xi 
(K_1,K_2,\ldots,K_N) \label{(8)} \eeq

The above relations become cumbersome analytically when the number of
coupling constants is large, but they can easily be verified for a few
one-coupling systems, to which we now turn.

\section{Examples}

We now consider a few specific examples in order to illustrate the
points made above. Perhaps the simplest example of a spin system which 
admits a closed-form, exact RG transformation is the one-dimensional 
Ising model with background and magnetic field terms, the reduced
Hamiltonian of which reads
\beq \CH = -K \sum_{(ij)} s_i s_j -h\sum_i s_i - \sum_i C \label{(9)}
\eeq
By a process of decimation, exact RG equations can be extracted
\cite{Yeo}. In order to be brief, we consider the case $C=h=0$, which
admits a one-parameter exact RG transformation defined by
\beq K'=\frac{1}{2} \log( \cosh (2 K)) \label{(10)} \eeq
The fixed point of (\ref{(10)}) is $K^*=0$. The correlation length is
\beq \xi (K) = \frac{1}{\log(\coth(K))} \label{(11)} \eeq
and using (\ref{(6)}) the beta function is easily seen to be
\beq \beta(K) =\frac{\log(\coth(K)) \coth(K)}{1-\coth^2(K)}
\label{(12)} \eeq
Then, considering the one-coupling version of (\ref{(4)}) it follows
that
\beq \frac{\log(\coth(K')) \coth(K')}{1-\coth^2(K')}= 
\frac{\log(\coth(K)) \coth(K) \tanh(2 K)}{1-\coth^2(K)} \label{(13)} \eeq
which is easily verified to hold by substituting $K'$ by its
expression (\ref{(10)}). Similar but messier considerations would hold
for the three coupling system (\ref{(9)}).

It is important to stress that the verification (\ref{(13)}) was not a
foregone conclusion before performing the calculation. To illustrate
this, consider the artificially modified RG flow
\beq K'=\frac{1}{2} \log( \cosh (2 K)) +\varepsilon K \label{(14)}
\eeq
where $\varepsilon$ is such that the iterates remain in the attraction
basin of the fixed point
$K^*=0$. When applied to (\ref{(4)}) an extra factor
appears and defining the deviation from consistency as
\beq \Delta = 1-\frac{\beta(K)}{\beta(K')} \frac{\pa K'}{\pa K}
\label{(15)} \eeq
we find
\beq \Delta = -\varepsilon \frac{\beta(K)}{\beta(K')} \label{(16)}
\eeq

This raises the obvious question of what happens when one considers
systems which do not have an exact RG transformation. The prototypical
example is the 2-D Ising model with no applied magnetic
field. Decimation procedures have been derived which reproduce the
observed behaviour qualitatively \cite{K&W} and to good numerical
accuracy \cite{Nied}. However, all these schemes are approximate, since
the Hamiltonians generated at each iteration step include next--to--next,
next--to--next--to--next... neighbour interactions; to keep the
equations manageable some truncation has to be performed. While this
truncation can be performed in such a way that the fundamental
requirement of Hamiltonian form invariance is respected, the RG flows
are no longer exact.

For simplicity, we consider a very crude decimation procedure on the
2D Ising, the Migdal-Kadanoff \cite{M&K} truncation. Its RG equation
reads
\beq K'= -\frac{1}{2} \log \frac{2 e^{-4K}}{1+e^{-8 K}} \label{(17)}
\eeq
Together with the correlation length \cite{length}
\beq \xi (K) = \frac{1}{|2 K - \log \tanh K|} \label{(18)} \eeq
we can check (\ref{(4)}). It turns out that the deviation $\Delta$
defined by (\ref{(15)}) evaluates to
\beq \Delta = 1 - 2 \tanh(K) \left(\frac{2 K + \log(\tanh
K)}{2+\frac{2}{\sinh(2 K)} }\right) \left(\frac{2 K'(K) + \log(\tanh
K'(K))}{2+\frac{2}{\sinh(2 K'(K))} }\right)^{-1} \eeq
which, when (\ref{(17)}) is substituted for $K'(K)$, is seen to be
non-zero. This, while far from unexpected, seems
to limit the applicability of the above considerations, since there
are few systems which admit an exact RG transformation. However, as
was seen from the artificially modified RG flow (\ref{(14)}), $\Delta$
can be viewed as a measure of how much a proposed RG flow differs from
an exact one, and this suggests that beta function consistency could
be used as an extra constraint in
numerical studies of RG schemes.

One further point that can be made is that the system (\ref{(4)}) can
be used to obtain the $t_i$ given the $\beta(K_i)$ (or, equivalently,
the correlation lengths). For instance, considering the
one-dimensional Ising beta function (\ref{(12)}), we have the
differential equation
\beq \frac{\pa K'(K)}{\pa K}= \frac{\log(\coth(K'(K)))
\coth(K'(K))}{1-\coth^2(K'(K))} \left(\frac{\log(\coth(K))
\coth(K)}{1-\coth^2(K)}\right)^{-1} \label{diff} \eeq
the solutions of which can be written in parametric form as
\beq \log\left(\frac{e^{2 K} +1}{e^{2 K} -1} \right) \log \left( \left( \frac{e^{2
K'(K)} +1}{e^{2 K'(K)} -1} \right) \right)^{-1} = C \label{sol} \eeq
where $C$ is a constant, related to the behaviour of the beta function
for a given point in coupling space. We see that (\ref{(4)}), while
generating a series of solutions which include the correct RG
equation, does not uniquely fix the solution. It does, however, provide a
strong hint that such a solution should exist, as well as suggesting its
general form.

\section{Conclusion, Discussion \& Outlook}

In conclusion, we have proposed a consistency relation between RG
flows and beta functions of spin systems which is obeyed by systems
having an exact RG transformation which leaves the form of the
Hamiltonian unchanged. We argued heuristically that a deviation from
such a consistency signals an inexact RG flow which, while quite
possibly accounting for the qualitative structure of the model, will
not produce a quantitative agreement with exact (or sufficient
accuracy numerical) results; the deviation from consistency could
provide one test of the ``goodness'' of proposed schemes.
We also proposed that the consistency relation, coupled with
information about the correlation lengths of a system, could be used
to suggest general forms of RG schemes. The limitations of this method
stem from the fact that some boundary conditions on the consistency
relations are left unspecified, but some sort of self-consistent
iterative procedure could perhaps be envisaged.

While written in the language of spin systems, it is felt that the
above considerations are of interest for QFTs as well, much in the
same way that the original arguments of Daamgard and Haagensen
\cite{D&H} have been. I hope to return to this matters in the future.

\end{document}